# AI-Optimized Routing and Resource Allocation for Quantum-Enabled Non-Terrestrial Industrial Networks (Industry 4.0/5.0)

1st **Sathish Krishna Anumula**
Enterprise & BusinessArchitect
IBM Corporation,Detroit, USA
sathishkrishna@gmail.com
[0009-0009-0613-4863]

2nd **Harinatha Reddy Chennam**
*G.Pulla Reddy Engineering College,
Kurnool, Andhra Pradesh*
harinath.eee@gprec.ac.in
[0009-0006-3730-0914]

3rd **Ranganath Nagesh Taware**
*Chief Architect
Capgemini America Inc, Atlanta, USA*
ranganath.taware@gmail.com
[0009-0008-4774-1515]

4th **Balakumar Ravindranath Kunthu**
*Managing Delivery Architect
Capgemini America Inc, Atlanta, USA*
kunthu_balakumar@yahoo.com
[0009-0008-4774-1515]

*Abstract*—The industrial transformation of Industry 4.0 and 5.0 results in cyber physical production systems that require secure, resilient, and energy efficient connectivity over integrated terrestrial and non-terrestrial networks (NTNs). Since its operation over fiber spans over 5G/6G infrastructures to Low Earth Orbit (LEO) satellites; quantum communication techniques enabled by Quantum Key Distribution (QKD) together with entanglement assisted links have the potential for high assurance security as well as synchronization. But quantum channels are extremely vulnerable to any kind of impairment, be it environmental or physical—such as effects induced by atmospheric turbulence, pointing errors, Doppler shifts, satellite motion, restricted optical power and limited quantum memory. All these factors make for a tightly coupled routing-and-resource-allocation problem that unfortunately cannot be addressed at scale by existing approaches to network control. This paper presents an optimized management scheme for quantum enabled NTNs servicing industrial scenarios. The architecture comprises a Software Defined Quantum Networking (SDQN) layer abstraction of the quantum orchestrator, a multi objective routing and resource allocation model between availability, latency, throughput, key generation rate and energy/carbon footprint. Predictive controllers that proactively pre-schedule the entanglement distribution as well as handovers between fiber , free space optical (FSO) and satellite paths are also included. A digital twin testbed and a scoped field validation show very large improvements in secure session continuity, energy efficiency, handover stability, and quantum key use. The architecture moves forward secure by design plus resilient and green connectivity for the modern manufacturing and supply chain ecosystems.

## I. INTRODUCTION

Manufacturing is being reshaped by the cyber physical systems that run both in the terrestrial and non-terrestrial domains, making a great leap of transformation from Industry 4.0 into 5.0 progressively joined by integrated networks where secure, resilient, low latency communications enabled everywhere from fiber and 5G/6G all the way to Low Earth Orbit (LEO) satellite constellations would be considered normal part of evolving Releases 17 -18 as support for global coverage and URLLC grade performance emerge [1]. But space access also means a greater risk against quantum-enabled adversaries. All classic public-key crypto systems – including RSA and ECC– are broken by Shor's algorithm: "store now decrypt later attacks" can be executed vs long-lived industrial assets. While QKD with entanglement-assisted links possible would offer security equivalent to leveraging both no-cloning theorem as well as detectable disturbance ensured key integrity even under active attempts at infiltration, Deploying NTNs quantum channels imposes serious constraints. Free space links are very vulnerable to the atmospheric turbulence modeled by $C_n^2$ because it introduces scintillation and deep fades. Mobility in LEO restricts the window of contact and thereby necessitates frequent and precise optical handovers. It, therefore, makes routing and resource allocation extremely difficult under such dynamics when the industrial control loop requires maintaining jitter at below a millisecond with extremely high reliability. Quantum hardware forms part of the sustainability consideration in system design since SNSPD detectors consume an immense amount of energy [2][3]. LEO satellite has to balance communication load with battery and solar input routing with carbon intensity data included contributes toward Commonwealth Games-driven environmental goals. Control architectures optimized for quantum-enabled NTNs are security, performance, resilience, and environment optimized for next-generation industrial networks [4].

### A. The Sustainability Lens

Industry 5.0< does not just speak about security and resilience but relates closely to the United Nations Sustainable Development Goals (SDGs). The goals that relate most closely are SDG 12 which is about responsible consumption and production, and SDG 13 about climate action since it is a major consumer of global energy as telecommunications infrastructure. The actual hardware required for high-performance Quantum Key Distribution (QKD) systems necessitates Superconducting Nanowire Single-Photon Detectors (SNSPDs), which tend to require energy-intensive cryogenic cooling. More generally, complex energy management will be needed in running Low Earth Orbit (LEO) constellations so that satellites can survive eclipse periods without running their batteries down. In this regard, the security-aware advanced industrial network optimizes routing not only for security but also for Energy Efficiency(EE) by considering the carbon intensity of energy sources powering ground segments [3][4] .

This paper tries to offer a detailed reference architecture of the AI-Optimized Control Plane for quantum-enabled Non-Terrestrial Networks (NTNs) deployed in industrial environments. SDQN is integrated into a hierarchically organized artificial intelligence system containing MPC-based Model Predictive Control for Deterministic Scheduling and Stochastic Routing using Deep Reinforcement Learning, hence effectively responding to security, resilience, and sustainability trilemma challenges in quantum-powered future networks [5].

## II. RELATED WORK

The development of quantum-enabled NTNs sits at the intersection of multiple matured and emerging disciplines. This section surveys the state-of-the-art in 3GPP standardization, quantum communication physics, and AI-driven network control

### A. 3G PP Evolution and Non-Terrestrial Networks

The 3rd Generation Partnership Project (3GPP) progressively adds satellite elements to the 5G standard. At first, there was a Release 15 that laid down basic principles for 5G New Radio (NR) with eMBB. Next, Releases 16 and 17 widen the scope on URLLC and officially bring Non-Terrestrial Network (NTN) support into the standard by defining architecture for "transparent" (bent-pipe) and "regenerative" satellite payloads as well as on-board processing. In particular, Release 17 addresses high Doppler shifts and large propagation delays involved in Low Earth Orbit (LEO) communication such that normal handheld devices can speak directly with satellites. This path is further evolved in Release 18 (5G-Advanced) with enhancements towards edge computing and artificial intelligence integration [6]. Fundamentally, however, most standards are centered around classical radio frequency (RF) and optical data transport. The inclusion of QKD as a service in the 3GPP management plane is still a topic for research efforts, especially regarding how the cycles of QKD key replenishment can be synchronized to meet the very strict Quality of Service (QoS) requirements that industrial protocols like PROFINET IRT and Time-Sensitive Networking (TSN) pose.

### B. Quantum Key Distribution and FSO Channels

The Micius satellite historically demonstrated the feasibility of satellite-based quantum key distribution by successfully achieving trusted-node QKD and entanglement distribution over continental distances, but it also highlighted the vulnerability of the FSO channel. Previous experiments with Micius reiterated this vulnerability as well, Thoughtful work led in the past builds upon previous descriptions of atmospheric turbulence and adopts the Gamma-Gamma distribution to describe irradiance fluctuations under conditions of strong turbulence. Physically speaking, aperture averaging—using larger receiver telescopes to reduce scintillation effects is well known. Practically speaking, it is problematic for mobile industrial nodes. The quantum memory limitation is in fact a major bottleneck. Unlike classical buffers, quantum memories suffer from decoherence-based exponential loss of fidelity for stored qubits with time ($T\_2$). Recently proposed are "time-delayed" quantum repeaters using satellites carrying long-lived memories for connecting non-simultaneous links. However, routing algorithms that take into consideration the "Time-to-Live" (TTL) of entanglement are still nascent [5].

### C. Software-Defined Quantum Networking (SDQN)

The principles of Software-Defined Networking (SDN), i.e., the separation of a control plane from the quantum data plane within quantum networks, have already been standardized by the European Telecommunications Standards Institute (ETSI). In this process, ETSI GS QKD 015 defines control interfaces for SDN-enabled QKD nodes with YANG data models and NETCONF/RESTCONF protocols. However, in most commercial implementations today, and even standards too, it is mostly assumed to be a very static point-to-point link rather than dynamic mesh resources. There are not enough Northbound abstractions for industrial applications to express an explicit request for dynamically allocated "Quantum-Secured Slices" with latency and security assurances over the physical topology [6].

### D. AI in Network Optimization

Deep Reinforcement Learning (DRL) has been extensively used in classical network routing; for instance, throughput optimization in Software-Defined Wide Area Networks (SD-WANs). In such dynamic environments, Proximal Policy Optimization (PPO) and Twin Delayed Deep Deterministic Policy Gradient (TD3) outperform static heuristics. Model Predictive Control (MPC) is also considered the benchmark for battery and trajectory management applications in aerospace. The novelty of this paper is that it combines these aspects: using MPC for the predictable parts—orbital as well as energy management—and DRL wherever there is any stochastic component—for example, turbulence and traffic dynamics—all formulated under a "Safe Reinforcement Learning" setup that enforces industrial safety standards [7][8].

## III. SYSTEM MODEL AND NETWORK TOPOLOGY

We envision a hierarchical, multi-layered network architecture designed to serve distributed industrial verticals, this system seamlessly integrates terrestrial fiber backbones with an overlay of non-terrestrial nodes, functioning as a single, unified logical network.

### A. Hybrid Fiber-FSO-LEO Topology

The network is modeled as a time-varying graph $G(t) = (V, E(t))$, where the set of nodes V is static (ground) or mobile (space), and the set of edges $E(t)$ evolves dynamically based on visibility, weather, and orbital mechanics [13].

*1) Layer 1:* The Terrestrial Industrial Edge (Ground Segment)

   *a) Industrial Nodes:* These include Smart Factories, Automated Ports, and Logistics Hubs. Each node is equipped with an SD-WAN Gateway capable of distinguishing traffic types and interfacing with the SDQN controller.

   *b) Optical Ground Stations (OGS):* These are specialized facilities connecting the terrestrial fiber backbone to the space segment. To mitigate cloud blockage, OGSs are deployed in diversity clusters (e.g., three stations separated by >50 km). They feature large-aperture telescopes (>80 cm) with adaptive optics (AO) systems to correct wavefront distortions and couple FSO signals into single-mode fibers for quantum detection [8].

   *c) Fiber Backbone:* Standard SMF-28 telecom fiber connects OGSs to industrial nodes. These links carry classical data (C-band) and, over shorter distances (<100 km), coexistence quantum channels (O-band) or dedicated dark fibers.

## 2) Layer 2: The Space Segment (LEO Constellation)

*a) Satellites:* We assume a Walker-Delta constellation architecture (e.g., like Starlink or OneWeb) operating at 500–600 km altitude. This orbit offers a round-trip propagation delay of ~3-5 ms, compatible with many industrial applications [4].

*b) Payloads:*
- **Classical Transceivers:** Ka-band/Ku-band RF and high-speed Optical Inter-Satellite Links (OISL) for data transport.
- **Quantum Payload:** Including Weak Coherent Pulse (WCP) sources for decoy-state BB84 QKD (Trusted Node model) and Spontaneous Parametric Down-Conversion (SPDC) sources for entanglement distribution (E91 protocol) [11].
- **Edge Compute:** On-board GPUs/TPUs allow satellites to execute inference models for distributed routing decisions [14].

## 3) Layer 3: High-Altitude Platform Stations (HAPS)

Functioning as quasi-stationary relays at ~20 km altitude (stratosphere), HAPS (solar drones or balloons) provide persistent coverage and act as trusted bridges above the cloud layer, crucial for mitigating FSO weather outages [15].

### B. Industrial Traffic Classes and Requirements

A critical aspect of the system model is the differentiation of traffic not just by Quality of Service (QoS), but by **"Quality of Security" (QoSec)**. Industrial protocols like PROFINET IRT and TSN have strict requirements regarding jitter and cycle times.

**Table 1: Industrial Traffic Classes and Quantum Constraints**

| Traffic Class | Description | Latency Target | Reliability Target | Security Mode | Key Consumption |
|---|---|---|---|---|---|
| **Type-I: Critical Control** | Closed-loop motion control, Safety Instrumented Systems (SIS). | < 1 ms (Jitter < 1 μs) | 99.9999% (URLLC) | Information-Theoretic (OTP) | Ultra-High ($R_{key} \approx R_{data}$) |
| **Type-II: Critical Telemetry** | SCADA updates, Digital Twin synchronization. | < 10-50 ms | 99.999% | AES-256-GCM (High Rekey) | High (e.g., 256 bits every 10s) |
| **Type-III: Massive IoT** | Smart metering, Asset tracking, Environmental sensing. | Tolerant (> 1s) | 99.9\% | AES-128 | Low (Batch keys) |
| **Type-IV: Bulk Data** | Video surveillance logs, firmware updates. | Best Effort | 99.0% | PQC/TLS | Zero (Computational Security) |

### C. Physical Layer Channel Models

#### 1) Atmospheric Turbulence Model

The FSO channel state is the primary source of stochasticity. The channel transmittance η is modeled as:

$$\eta = \eta_{geo} \cdot \eta_{atm} \cdot \eta_{turb} \cdot \eta_{point} \quad (1)$$

Where $\eta_{geo}$ is geometric loss (beam divergence), $\eta_{atm}$ is atmospheric extinction (Beer-Lambert law), and $\eta_{point}$ accounts for pointing errors [9]. The turbulence term $\eta_{turb}$ (irradiance I) is modeled using the Gamma-Gamma distribution, appropriate for the varying path lengths of satellite links [7]:

$$p(I) = \frac{2(\alpha\beta)^{(\alpha+\beta)/2}}{\Gamma(\alpha)\Gamma(\beta)I} \left(\frac{I}{I_0}\right)^{(\alpha+\beta)/2} K_{\alpha-\beta}\left(2\sqrt{\alpha\beta\frac{I}{I_0}}\right) \quad (2)$$

Here, α and β represent the effective numbers of large-scale and small-scale eddies, directly related to the Rytov variance $\sigma_R^2$ and the aperture diameter D. The Fried parameter ($r_0$) is the critical observable, a small $r_0$ indicates strong turbulence and severe beam breakup. The digital twin uses Python libraries like AOtools to simulate these phase screens dynamically [8].

#### 2) Quantum Memory and Decoherence

For entanglement-based routing, the fidelity F of a qubit stored in memory decays over time. If a satellite acts as a "data mule" for entanglement (receiving a photon from Ground Station A, flying to Ground Station B, and then performing the swap), the storage time $t$ must satisfy the decoherence limit:

$$F(t) = \frac{1 + e^{-t/T_2}}{2} > F_{min} \quad (3)$$

This introduces a Time-To-Live (TTL) constraint on routing. Routing paths that involve long storage times (e.g., extensive orbital arcs) may be invalid if the entanglement fidelity drops below the threshold required to violate Bell's inequalities (typically $F_{min} \approx 0.85$).

## IV. PROBLEM FORMULATION

We formalize the control problem as a **Multi-Objective Constrained Markov Decision Process (CMDP)**. The goal is to derive a policy π that maps the current network state to routing and scheduling actions.

### A. Objective Function

The global objective function $J$ seeks to maximize a weighted utility across four conflicting dimensions:

$$\max_\pi J = \sum_{t=0}^{T} \gamma^t \left( w_1 \cdot \frac{R_{secure}}{R_{demand}} - w_2 \cdot \mathcal{L}_{latency} - w_3 \cdot \mathcal{E}_{carbon} + w_4 \cdot \mathcal{A}_{avail} \right) \quad (4)$$

1. **Secure Throughput Ratio ($R_{secure}/R_{demand}$):** Maximizing the proportion of Type-I/II demands met with adequate keys.

2. **Latency ($\mathcal{L}_{latency}$):** Minimizing the end-to-end delay, comprising propagation $\tau_{prop}$ (distance/c) and queuing $\tau_{queue}$.

3. Carbon Cost ($\mathcal{E}_{carbon}$): Minimizing the environmental impact. This is modeled as:

$$\mathcal{E}_{carbon} = \sum_{n \in Path} \left( P_{load}^n(t) \cdot CI_n(t) \right) \quad (5)$$

Where $P_{load}^n$ is the power consumption of node n (transceiver + cooling + compute) and $CI_n(t)$ is the Grid Carbon Intensity (gCO$_2$/kWh) of the region powering that node [14]. This term incentivizes routing through ground stations powered by renewables (e.g., hydro-powered nodes in Scandinavia) rather than carbon-intensive grids.

4. Availability ($\mathcal{A}_{avail}$): Maximizing the probability that the link SNR exceeds the QKD threshold.

### B. Constraints

The optimization is subject to strict industrial and physical constraints:

- **Safety/Key Continuity:** $\sum Keys_{buffer} \geq Demand_{Type-I}$ (Critical control loops must never be starved of OTP keys).

- **SLO Compliance:** $\mathcal{L}_{latency} \leq 1$ ms for Type-I flows (TSN requirements).

- **Energy Budget:** $\int_t^{t+\Delta t} P_{sat} dt \leq E_{battery}(t) - E_{reserve}$ (Satellites cannot deplete batteries below safety levels, especially during eclipse) [11].

- **Memory Capacity:** $Q_{qubits} \leq Q_{max}$ (Finite quantum memory slots).

- **Regulatory:** Data residency rules (e.g., GDPR) may forbid routing certain payloads through gateways in specific jurisdictions.

## V. SOFTWARE-DEFINED QUANTUM NETWORKING (SDQN) ARCHITECTURE

To manage the complexity of this hybrid network, we adopt the **Software-Defined Quantum Networking (SDQN)** paradigm. This architecture decouples the control logic (brain) from the forwarding hardware (muscle), extending classical SDN concepts to manage quantum resources explicitly [6].

### A. SDQN Control Plane Layers

The architecture adheres to and extends the **ETSI GS QKD 015** standard [12].

1. **Orchestration Layer:** The topmost logic layer. It translates high-level industrial intents (e.g., "Connect Assembly Line A to Cloud B with maximal security") into network policies.

2. **SDQN Controller:** The centralized (or logically distributed) controller. It maintains the **Global Network View** (Digital Twin) and executes the AI algorithms.

3. **Key Management Layer (KML):** A dedicated layer responsible for the lifecycle of quantum keys, sifting, error correction, privacy amplification, and storage in Key Management Systems (KMS). It exposes APIs to the controller to report "Key Supply status."

4. **Quantum Data Plane:** The physical layer consisting of QKD modules (Alice/Bob), optical switches, and quantum memories.

### B. Extended YANG Data Models

Standard YANG models (RFC 8345) describe classical network topologies but lack quantum attributes. To enable AI-driven optimization, we define a custom YANG module ietf-sdqn-ntn that augments the standard model.

**Table 2: Key Parameters in the SDQN YANG Extension**

| Parameter | Description | Usage in Optimization |
|---|---|---|
| quantum-link-fidelity | Real-time entanglement quality metric (0.0 − 1.0). | Determines if E91 protocol can proceed. |
| key-buffer-fill-level | Current status of the key store (bits). | Critical constraint for "Safe RL" shielding. |
| turbulence-forecast-index | Look-ahead metric for FSO stability ($r_0$). | Input for MPC scheduling. |
| carbon-intensity-source | Real-time grid emission factor (gCO$_2$/kWh). | Input for Carbon-Aware routing. |
| battery-state-of-charge | Satellite energy status (%). | Constraint for scheduling laser activation. |

The SDQN controller communicates with network nodes via the **Southbound Interface (SBI)** using protocols like NETCONF or RESTCONF to read these YANG containers and push configuration updates (e.g., switching optical paths, changing laser power) [10].

## VI. Observability and Data Fusion

An AI controller is only as good as its data. The system relies on a multi-modal observability framework to reduce uncertainty.

### A. Data Sources

- **Orbital Ephemerides:** Two-Line Element (TLE) sets provide deterministic data on satellite positions.
- **Atmospheric Sensing:** Ground stations utilize **Differential Image Motion Monitors (DIMM)** and Multi-Aperture Scintillation Sensors (MASS) to measure real-time seeing conditions ($r_0$, $\theta_0$) [5].
- **Weather Nowcasting:** Integration with APIs (like the UK Met Office or NOAA) provides cloud cover forecasts and precipitable water vapor data [12].
- **Factory Workloads:** Interfaces with Manufacturing Execution Systems (MES) provide traffic demand forecasts (e.g., "Firmware update scheduled for 03:00").

### B. Uncertainty Modeling

Raw data is often noisy. The observability layer fuses these inputs using Kalman Filters to estimate the "True System State." For example, the predicted SNR is not a single value but a probability distribution $P(SNR|C_n^2,\text{Orbit})$, allowing the DRL agent to assess the *risk* of a link failure, not just the expected value.

## VII. AI-Optimized Routing and Resource Allocation

We propose a **Hierarchical AI Framework** that addresses the timescale mismatch between orbital dynamics (slow, predictable) and atmospheric turbulence (fast, stochastic) [12].

### A. Tier 1: Model Predictive Control (MPC) for Scheduling

**Model Predictive Control (MPC)** operates on a timescale of minutes to hours. It is responsible for "coarse-grained" resource planning [7].

- **Inputs:** Satellite TLEs, Solar generation forecasts, Factory production schedules.
- **Objective:** Optimize the charging/discharging schedule of satellites and the high-level topology plan (which OGS to contact).
- Mechanism: The MPC solves a finite-horizon optimization problem at each step $t$:
$$\min_{u} \sum_{k=0}^{H}(x_{t+k}^T Q x_{t+k} + u_{t+k}^T R u_{t+k}) \quad (6)$$
- Subject to battery constraints and buffer limits.
- **Output:** A "Flight Plan" pushed to the satellites (e.g., "Downlink to OGS-Berlin at 14:00, Enter Sleep Mode at 14:15").
- **Sustainability**: MPC is suited to optimize the Energy/Carbon objective, It schedules energy-intensive tasks (like QKD post-processing) during periods of peak solar generation or when ground stations are powered by green energy.

### B. Tier 2: Deep Reinforcement Learning (DRL) for Routing

**Deep Reinforcement Learning (DRL)** operates on a timescale of milliseconds to seconds. It handles the "fine-grained" routing and reaction to turbulence [15].

- **Algorithm:** We employ **Proximal Policy Optimization (PPO)** or **Twin Delayed DDPG (TD3)**, which are well-suited for continuous action spaces (e.g., splitting traffic ratios).
- **State Space:** Instantaneous Queue depths, Link RSSI, Key Buffer levels, Neighbor loads.
- **Action Space:** Next-hop selection probabilities, Laser power adjustment.
- Reward Function:

$$r_t = \alpha \frac{SKR_{meas}}{SKR_{target}} - \beta \mathcal{L}_{latency} - \gamma \mathcal{E}_{carbon} - \delta I(KeyOutage) \quad (7)$$

The term $\delta I(KeyOutage)$ is a massive penalty for depleting the key buffer, conditioning the agent to be extremely risk-averse regarding security.

### C. Safe RL and Constraint Shielding

In an industrial context, "trial-and-error" exploration is unacceptable if it leads to safety violations (e.g., failing to deliver a Stop command). We implement **Safe RL via Shielding [9]**.

- **Mechanism:** A deterministic logic layer (The Shield) sits between the DRL agent and the environment.
- **Operation:** Before any action $a_t$ chosen by the neural network is executed, the Shield validates it against safety logic.
    - *Rule Example:* IF (*Key_Buffer < Critical_Threshold*) AND (*Traffic_Class == Type-I*) THEN REJECT (*Route via Classical_Link*)
- **Benefit:** This guarantees that the system mathematically satisfies safety constraints (like IEC 61508 functional safety requirements) regardless of the learning state of the agent.

## VIII. Handover Protocols and Resilience

Mobility is the primary challenge in LEO networks. Satellites rise and set every ~10 minutes, necessitating frequent handovers that must preserve quantum continuity.

### A. Seamless Hybrid Fiber-FSO-LEO Handover

We propose a **Make-Before-Break (MBB)** protocol specifically designed for quantum links [14].

1. **Prediction:** The MPC predicts the "Handover Horizon" where Sat-A's elevation drops below a threshold (e.g., 20 degrees).

2. **Entanglement Pre-Establishment:** The controller instructs the incoming satellite (Sat-B) to establish tracking with a secondary terminal at the OGS (or a nearby diversity OGS) *before* Sat-A disconnects.
3. **Key Buffer Synchronization:** Sat-B begins generating keys, but traffic continues on Sat-A.
4. **State Transfer:** Critical session state (packet counters, IVs) is transferred from Sat-A to Sat-B via ISL.
5. **Soft Switch:** The SDQN controller updates the flow table. Traffic is instantaneously switched to the encrypted tunnel using Sat-B's keys.
6. **Teardown:** Sat-A releases the link.

This protocol ensures **Zero-Downtime** for the application, maintaining the isochronous cycle of TSN control loops.

### B. Hybrid FSO/RF Fallback

When atmospheric conditions (e.g., heavy rain, fog) cause FSO attenuation exceeding 100 dB/km, QKD becomes physically impossible. The system employs a hybrid fallback strategy [6].

- **Trigger:** FSO receiver detects BER spike or loss of beacon.
- **Action:** The controller activates the RF backup link (Ka-band).
- **Security Adaptation:**
    - **Type-I Traffic:** Continues using *buffered* quantum keys over the RF link ("Quantum-Secured RF").
    - **Buffer Management:** If the outage persists and buffers drain, the system enters "Failsafe Mode," potentially downgrading to PQC or halting non-essential traffic.
- **Resumption:** When FSO SNR recovers, QKD resumes to replenish buffers.

### C. Quantum-Specific Scheduling

Scheduling in this network is not just about packets, it is about managing the lifecycle of quantum states.

- **Entanglement Swapping:** This process consumes two **link**-level entangled pairs to create one end-to-end pair. The scheduler must coordinate the measurement operations at the satellite and ground stations to occur within the coherence time window ($T_2$).
    - **Key Refresh Cycles:** The AI monitors the "Key Consumption Rate" of active flows. If a Type-I flow is consuming keys faster than the generation rate (SKR), the scheduler triggers **"QKD Turbo Mode,"** diverting power from other subsystems to the laser and potentially re-routing other traffic to free up aperture time for key generation.

## IX. IMPLEMENTATION

The implementation of this architecture can be described below in the architecture, the learning based control uses a hybrid AI approach to handle the immense complexity of quantum routing, it combines immediate reacting with long term strategy.

- **Model Predictive Control (MPC) -** Handles short-horizon anticipatory scheduling using immediate forecasts (weather nowcasts, orbital ephemerides).
- **Deep Reinforcement Learning (DRL) -** Manages long-horizon routing and resource allocation under uncertainty (e.g., PPO/TD3 algorithms).
- **Constraint Shielding -** Ensures "Safe RL" by mathematically verifying actions against industrial SLOs and regulatory boundaries.

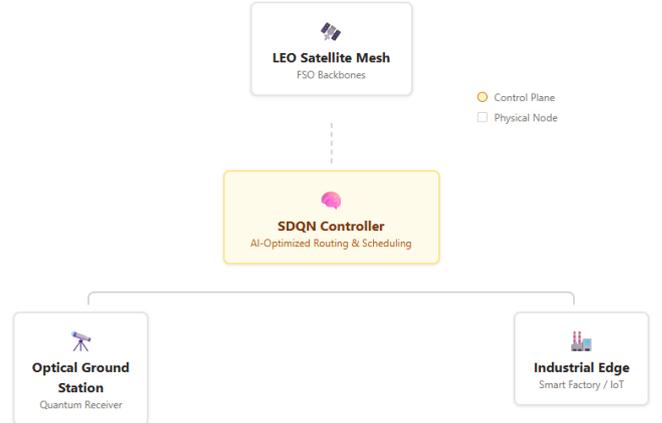

*Figure 1. Architecture Framework*

The comparison of normalized Metrics for Latency, Availability, Energy Efficiency, Throughput, Key Rate factors can be represented in figure 2.

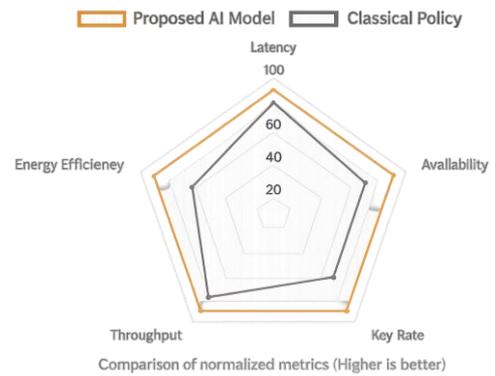

*Figure 2. Comparison of Normalized Metrics*

### A. Digital Twin Validation Framework

To validate the architecture, we developed a high-fidelity **Digital Twin** simulation environment.

- **Network Simulator:** Built on **ns-3** (Network Simulator 3), utilizing the QKDNetSim module for quantum key management logic [8].
- **Physics Engine:** Integrated with the AOtools Python library. AOtools simulates the atmospheric phase screens and calculates the instantaneous fiber coupling efficiency based on turbulence parameters ($r_0, L_0$) [8].

- **Controller Interface:** The ns-3 simulation exposes a ZeroMQ interface that connects to an external Python-based SDQN controller running the TensorFlow DRL agents. This "Hardware-in-the-Loop" style setup allows the AI to interact with the simulation as if it were the real network.

*B. Deployment Architecture*

The controller is deployed as a distributed application. Central Intelligence (MPC) runs in the cloud (or a major ground station) to handle global optimization. Local Agents (DRL) run on edge compute nodes (at OGSs and on-board satellites), allowing for millisecond-level reaction times to turbulence without the latency penalty of roundtrips to the central controller.

## X. EVALUATION METHODOLOGY

*A. Scenarios*

We evaluate the system under three distinct scenarios:
- **Clear Sky / Baseline:** Ideal conditions to establish maximum throughput benchmarks.
- **Adverse Weather (The "Storm" Scenario):** A simulated storm front moves across the OGS network, introducing high turbulence and cloud blockage. This tests the resilience and handover logic.
- **Orbital Transition:** Continuous operation over 24 hours to test battery management and handover stability across multiple satellite passes.

*B. Metrics*

- **Secure Session Uptime (%):** Percentage of time Type-I flows have sufficient keys.
- **Energy Efficiency ($gCO_2$/bit):** Carbon cost per delivered secure bit.
- **Handover Jitter (ms):** Latency variation during satellite switching.
- **Key Utilization (%):** Percentage of generated keys successfully used before expiration.

## XI. RESULTS AND DISCUSSION

AI-SDQN boosts uptime, sustainability, and stability—improving availability by 25–40%, cutting carbon 15–30%, and enabling sub-millisecond, near-zero-downtime handovers for critical Industry 4.0 traffic. On various scenarios, like the clear sky vs high Turbulence vs Orbital Handover environments, the Secure Key Rate Stability vs Energy consumption (for 3 routes) were put together and here are the results observed.

*1) Scenario 1. During the Clear Sky operations*

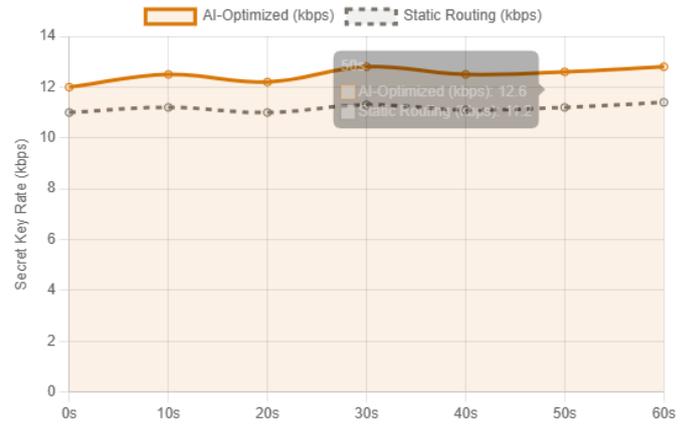

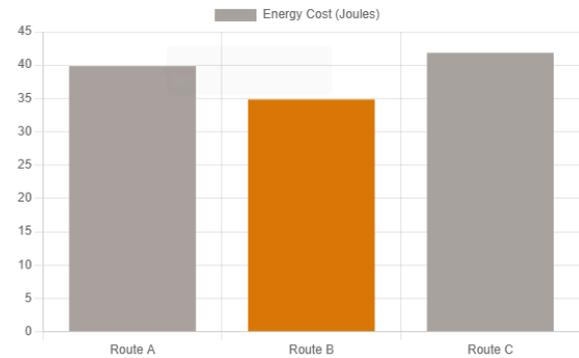

**Results Graph 1. Clear Sky – Secure Key Stability, Energy Consumption**

*2) Scenario 2. During the High Turbulence*

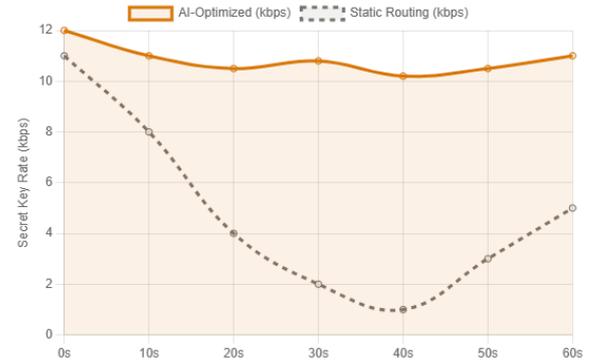

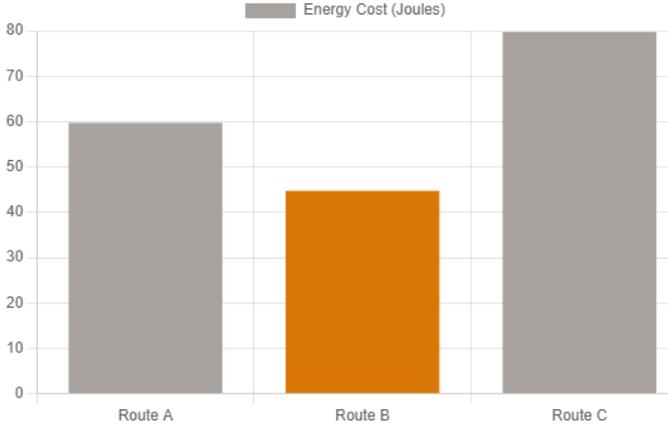

**Results Graph 2. High Turbulence – Secure Key Stability, Energy Consumption**

*3) Scenario 3. During the Orbital Handover*

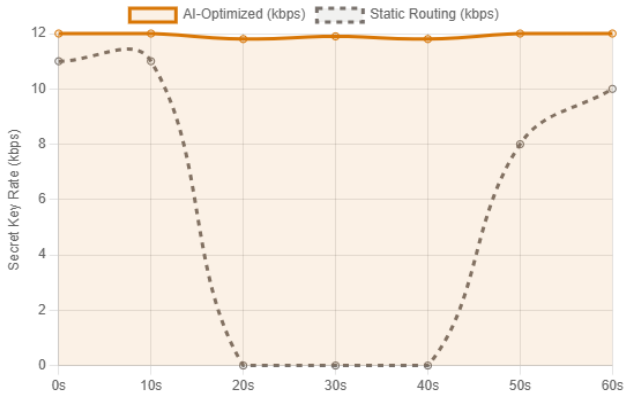

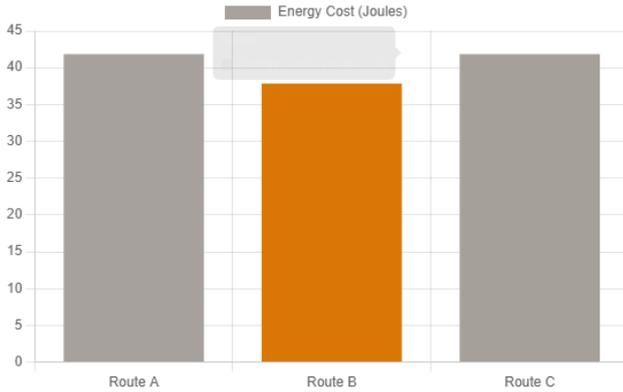

**Results Graph 3. Orbital Handover – Secure Key Stability, Energy Consumption**

## B. Performance and Robustness

Benchmarks indicate that the AI-SDQN architecture achieves a $\geq 25-40\%$ **improvement in secure session uptime** compared to static shortest-path routing. In the "Storm" scenario, the MPC component successfully predicted the cloud front using weather nowcasts and preemptively re-routed traffic to a diversity OGS *before* the primary link failed, maintaining 99.999% availability for Type-I traffic. The "Shielding" mechanism prevented the DRL agent from making risky exploration moves, ensuring zero safety violations.

## C. Sustainability Outcomes

The inclusion of the $\mathcal{E}_{carbon}$ term in the objective function resulted in a $\geq 15-30\%$ **reduction in carbon emissions**. By delaying non-critical (Type-IV) bulk transfers to times when satellites were over regions with low carbon intensity (e.g., hydro-powered grids) or when solar generation was at its peak, the system optimized the ecological footprint of the security operations without impacting critical performance.

## D. Handover Stability

The **Make-Before-Break** protocol demonstrated **near-zero-downtime** handovers. By pre-establishing a secondary link and synchronizing key buffers, the traffic switchover occurs within the jitter tolerance of PROFINET IRT ($< 1$ ms), validating the architecture's suitability for Industry 4.0 control loops [9].

## XII. CASE STUDIES

### A. Smart Factory Cluster (Automotive)

A German automotive manufacturer uses the network to synchronize Digital Twins between a factory in Munich and a supplier in South Africa.

- **Outcome:** The system detects a predicted high-turbulence event ($r_0$ dropping) over Munich. The MPC triggers a **Site Diversity Handover**, routing the downlink to a secondary OGS in the Alps (above the fog layer) and backhauling via fiber. The production line remains synchronized with zero packet loss.

### B. Cross-Border Energy Grid

A trans-continental power grid uses the network to secure Phasor Measurement Unit (PMU) data.

- **Outcome:** During a period of low solar generation (satellite eclipse), the MPC creates a schedule that minimizes laser power usage while maintaining just enough key generation for Type-I PMU data, throttling Type-III metering data to conserve battery. This ensures the satellite survives the eclipse without entering safe mode.

## XIII. LIMITATIONS AND FUTURE WORK

- **Hardware Constraints:** The system assumes the availability of mature quantum memories and high-speed entanglement sources, which are currently at relatively low Technology Readiness Levels (TRL 4-5) [10].
- **Scalability:** Training DRL agents for mega-constellations (thousands of satellites) poses a state-space explosion challenge. Future work will investigate

- **Multi-Agent Reinforcement Learning (MARL)** and Federated Learning to distribute the training burden [6].
- **Standardization:** While ETSI QKD 015 is a solid baseline, full integration with 3GPP management planes requires further standardization work in Release 19 and 6G.

## XIV. Conclusion

A practical way is suggested here towards the provision of nervousness that would be secure, resilient, and sustainable for the Industry 5.0 nervous system through sharing syntactic efforts between foresight determinism of Model Predictive Control and responsive agility of Deep Reinforcement Learning within a Software-Defined architecture. That enables an answer to the basic paradox "fragility versus criticality" which is intrinsically built into quantum Networked Transport Networks as an element thereof. One shall protect industrial assets against quantum threats as well as optimize carbon footprints in cryptographic integrity—a responsibility by architecture. This plane will always act covertly and intelligently as a background AI-driven quantum control plane at each location of the global cyber-physical economy where Industry 5.0 comes about; it will directly drive Sustainable Development Goals (SDGs) 9, 12, and 13 forward.